# Uniaxial "nematic-like" electronic structure and Fermi surface of untwinned $CaFe_2As_2$

Q. Wang,[1] Z. Sun,[1] E. Rotenberg,[2] F. Ronning,[3] E. D. Bauer,[3] H. Lin,[4] R. S. Markiewicz,[4] M. Lindroos,[4] B. Barbiellini,[4] A. Bansil,[4] and D. S. Dessau[1]

[1]Department of Physics, University of Colorado, Boulder, CO 80309-0390, USA.
[2]Advanced Light Source, Lawrence Berkeley National Laboratory, Berkeley, California 94720, USA.
[3]Los Alamos National Laboratory, Los Alamos, NM 87545, USA.
[4]Physics Department, Northeastern University, Boston, MA 02115, USA.

Obtaining the electronic structure of the newly discovered iron-based superconductors is the key to understanding the mechanism of their high-temperature superconductivity. We used angle-resolved photoemission spectroscopy (ARPES) to make direct measurements of the electronic structure and Fermi surface (FS) of the untwinned uniaxial state of $CaFe_2As_2$, the parent compound of iron-based superconductors. We observed unequal dispersions and FS geometries along the orthogonal Fe-Fe bond directions. More importantly, unidirectional straight and flat FS segments are observed near the zone center, which indicates the existence of a unidirectional nematic charge density wave order, strengthening the case for a quantum electronic liquid crystalline "nematic" phase. Further, the doping dependence extrapolates to a possible quantum critical point of the disappearance of this order in the heavily overdoped regime of these materials.

The cuprates (1) and the newly discovered iron pnictides (2) as two classes of unconventional superconductors share a lot in common: both are layered systems, transition metal d-electrons are important in both systems, the non-superconducting parent compounds of both systems exhibit anti-ferromagnetic (AFM) order and both systems become superconducting by a certain level of electron- or hole- doping. Although, unlike the cuprates whose parent state is a Mott insulator, the parent state of iron pnictides is metallic, the transport (2) and optical (3) measurements on iron pnictides show a very large resistivity and a small Drude weight. This indicates that the iron pnictides may not be conventional metals and the electron correlations could also play a crucial role as they do in the cuprates. It has been proposed that in strongly correlated electron systems, the strong electron correlations could lead the system into exotic quantum electronic liquid crystalline phases with translational and rotational symmetries broken ("stripes" or smectic phase) or only rotational symmetry broken (nematic phase) (4). Recent experimental studies provided growing evidence that the electron nematic phase exists in under-doped cuprates (5-9). As for iron pnictides, the latest scan tunneling microscope (STM) (10) and transport (11) studies also indicate a possible nematic phase. If it is true that the nematic phase exists in both systems, it will suggest another important common point between cuprates and iron pnictides, especially as to a potential quantum critical point when such order disappears with doping.

Fig. 1(a) shows a typical phase diagram for $AFe_2As_2$ type (A=Ba, Sr, or Ca) iron pnictides. The inset is the schematic of the in-plane crystal structure in the magnetic phase. The AFM order is characterized by its collinear properties: an AFM arrangement of ferromagnetic (FM) chains. This magnetic phase transition is coupled to a weak (~1%) tetragonal to orthorhombic crystal structure phase transition, which happens at (un-doped compound) or slightly above (doped compound) the magnetic phase transition temperature (12) as shown in the phase diagram. Considering the importance of electron correlations, the local-spin-moment picture (13-15) has been proposed to explain the orthorhombic distortion and the magnetism in the iron pnictide system. The recent STM (10), transport (11) and inelastic neutron scattering (16) measurements all indicate a 2-fold ($C_2$) symmetry. This anisotropic property of the system could be explained by the additional Ising degree of freedom (17) or the Pomeranchuk instability from functional renormalization group theory (18). Moreover, the nematic orbital ordering has been proposed by several theoretical groups (19-23) to explain the phase transition and breaking of 4-fold symmetry.

To examine the validity of the above theories and the authenticity of the existence of the nematic phase in iron pnictides, a comprehensive understanding of the electronic structure of the iron pnictide system is badly needed. ARPES, as the method which allows direct access to the electron dispersion and spectral function in condensed matter systems, is one of the best methods to study the electronic structure of the strongly correlated electron system such as cuprates (24, 25) and has also been performed on iron pnictides (26-30). One difficulty of ARPES measurements on iron pnictides is the formation of a pattern of twin domains (31). Generally, the domain size is smaller than the photon beam size in the experiment, so that most of the current ARPES studies on iron pnictides presented the domain averaged result and the intrinsic electronic structure is still missing. In our present study, we measured the high quality $CaFe_2As_2$ crystals with a very small photon beam size. This, together with the relatively large single domain area on the crystal surfaces, allowed us to make measurements of monodomain regions of the cleaved sample surfaces. The ability to deconvolve the twinned structure combined with our detailed polarization dependent studies, $k_z$ dependent studies, and Local-Density

Approximation (LDA) calculations, enabled us to make the most comprehensive analysis of the electronic structure of a pnictide to date.

In the inset of fig. 1(a) within the AFM-orthorhombic state, the AFM spin ordering happens along the x-direction in the crystal coordination while the FM spin ordering happens along the y-direction. Fig. 1(b) shows an overview of the experimental FS of $CaFe_2As_2$ in the AFM-orthorhombic state taken with 80eV photons. There is a well known hole pocket at the zone center highlighted in blue as well as an electron pocket at the zone corners highlighted in green, with the AFM nesting vector $\vec{q}_0$ joining them. This paper shows that calling these "pockets" is actually a misrepresentation, as there is much detail within each of them and they are far from being simple hole or electron pockets. In particular, we see that the internal structure within the "pockets" does not show a 4-fold symmetry: there are closed small Fermi pockets along one Fe-Fe bond direction while they are absent along the orthogonal direction. By comparing with the LDA calculation (details discussed later), we assign the direction that contains the small Fermi pockets to the y-direction (FM direction) and the orthogonal direction to the x-direction (AFM direction). Then the $\vec{q}_0$ in the plot indicates the collinear AFM ordering vector. Panels (c) to (g) show the FS and intensity maps at different binding energies. In this case, we rotated the sample's in-plane angle by 90 degrees and these small pockets rotate with the sample as it is rotated, indicating that they are an intrinsic aspect of the electronic structure and are not, for example, a result of the photon polarization. Furthermore, along the y-direction, the small Fermi pockets are electron-like while along the x-direction there are only hole-like features, i.e., when going to deeper binding energy the spectral weight spreads further away from the zone center. Hence, the plots shown in fig. 1 present a clear uniaxial electronic structure of $CaFe_2As_2$ in the AFM-orthorhombic phase.

If one takes a closer look at the small Fermi pockets along the y-direction, two long parallel straight FS segments can be found which indicates an incommensurate FS nesting as indicated by $\vec{q}_1$ in the fig. 1(b). Figs. 2(a) and (b) show the FS's in the AFM-orthorhombic phase taken with 80eV and 99eV photons, respectively. Here we utilize the tetragonal Brillouin zone (BZ) so as to be able to utilize standard orbital symmetry labels. The photon energy dependent studies (shown in panels (g) and (h)) indicate that the 80eV/99eV photons probe the electronic structure in the Z/Γ plane in the momentum space. Hence the FS cuts at Z and Γ planes both show the incommensurate FS nesting with approximately the same nesting vector $\vec{q}_1$=0.33(-π/a, π/a).

Figs. 2(c)-(d) and (e)-(f) show the dispersions along the AFM (yellow cut) and FM directions (blue cut) taken with 80eV and 99eV photons, respectively. Along ΓX/ZX' direction, there are three bands with different Fermi crossings near the zone center while along ΓY/ZY' direction, there is a clear band folding with both 80eV and 99eV data. Figs. 2(g) and 2(h) show the $k_z$ dispersions of the Fermi crossings along ΓX/ZX' and ΓY/ZY' directions. Consistent with the uniaxial electronic structure, panels (g) and (h) exhibit different $k_z$ dispersions. More importantly, along ΓY/ZY' direction, where the incommensurate FS nesting happens, the Fermi crossings of the nesting bands have minimal $k_z$ variation, as indicated by the solid blue lines in panel (h). This result confirms that the incommensurate FS nesting observed along ΓY/ZY' direction (FM direction) persists through the whole momentum space with an essentially unchanged nesting vector $\vec{q}_1$=0.33(-π/a, π/a).

To further understand the electronic structure of $CaFe_2As_2$, a polarization-dependent ARPES study using three different experimental geometries (fig. 3 rows a, b and c) has been performed. This study allows us to emphasize or de-emphasize different states according to their orbital symmetries – for example we turn on the flat portions of FS near the zone center in panel (c2) but turn these off and turn on complementary portions of the FS in panel (b2). Experimental dispersions along high symmetry cuts (white dashed lines in panels (a2), (b2) and (c2)) are shown in (a3), (b3) and (c3), while (a4), (b4) and (c4) show the extracted dispersions along these cuts as well as the allowed orbital symmetries. Here we note that for setup (a1) and (b1), the small beam size gives us a chance to probe the un-twinned electronic structure along different direction in the crystal while for setup (c1) the relatively larger beam spot always gives out the twinned result in the experiment. So for setup (c1), both x and y directions in the crystal could be aligned along the analyzer's slit direction at the same time due to the domain averaging effect.

Fig. 4 shows a compilation of the FS data, dispersion data, and orbital data, as well as a comparison to our LDA calculations. Figs. 4(a1) and 4(a2) are the experimentally extracted dispersions and FS with symmetry information color-coded. Here we ignore the possible $x^2-y^2$ and $z^2$ symmetry component for simplification, consistent with the band calculation which indicates that the near $E_F$ states are dominantly the xy and xz/yz states (32). Figs. 4(b1) and 4(b2) are raw data taken with mixed polarizations so as to show all symmetry states. Figs. 4(c1) and (c2) present the FS and dispersions along the high symmetry directions (at $k_z=2\pi/c$) based on our LDA calculations (based on the KKR methodology for complex crystals (33)).

Here we found that with a bandwidth renormalization factor of 2.5, the LDA calculations with magnetic moment of 0.19 $\mu_B$ give us the best match to the experimental dispersions and FS. The renormalization factor of 2.5 which is similar to what has been determined from other experiments (28, 29) is one of the aspects that indicate the importance of the electron correlations in the pnictides. To further improve the agreement between the theory and experiment, we shift the two calculated bands with yz symmetry (green dashed lines) up by 0.1eV (green solid lines) to match the bands $\gamma 1$ and $\gamma 2$ in fig. 4(a2). After this modification, the overall band calculation matches the experimentally determined dispersions reasonably well. Here we note that the up-shifting of the bands with specific orbital property may be a sign of the developing of the orbital ordering, which has been advanced by recent theoretical studies (19-23). Recent LDA+DMFT calculations (34) reported that the correlation effect in pnictides could lead to an orbital dependent band shifting compared to the pure LDA calculation, which is qualitatively consistent with our observation. This indicates that the electron correlation effect missing in the LDA calculation may account for this up-shifting.

Beside the similarities between the experimental and theoretical dispersions and FS after the optimizations mentioned above, there are also some important differences as well. In the experiment, we only observe one electron pocket ($\alpha 1$) at X' point but nothing at Y' point while the calculation indicates a relatively complex dispersion near the zone corner. The symmetry properties obtained from the experiment are also not fully consistent with the calculation: along ZY' direction, the FS contains two electron pockets which are formed by the bands with different symmetry properties ($\gamma 1$, $\gamma 2$ with xy/xz symmetry and β1, β2 with yz/$x^2$-$y^2$/$z^2$ symmetry) while the calculation indicates that both bands should be mainly of xy symmetry for one of these two pockets.

Accompanying this discrepancy in symmetry properties, a more interesting feature is found for the dispersion along the ZY' direction: there is no gap opened between the bands that form the electron pockets. This indicates that the electron pockets along ZY' direction are not formed in the traditional spin-density-wave manner but more like a Dirac cone structure instead, which is consistent with the orthogonal symmetry properties we have observed for the bands that make up these electron pockets. We note that this observation is consistent with the "nodal spin density wave" picture proposed by Ran *et al.* (35), in which a symmetry enforced band degeneracy at high symmetry points causes the existences of the nodes in the SDW gap function and leads to a Dirac-cone-like band structure. Furthermore, recent ARPES studies on $BaFe_2As_2$ (36) also suggest that band γ1 and γ2 along ZX' direction could form a Dirac-cone-like feature and present a tiny pocket at Fermi level, though they have not shown the orthogonal symmetry of the relevant bands. In our case, the un-closed FS segments along ZX' direction in our proposed FS also may be due to the formation of this Dirac cone but the current experimental condition does not allow us to fully resolve this tiny feature to get a conclusive result.

Finally, we focus on the unidirectional incommensurate FS nesting observed in the experiment. The very flat sections of FS which give rise to the nesting are not present in the theoretical calculations we presented here, which include the effects of the uniaxial spin order and which have been optimized for best agreement with the experimental data (37). Furthermore the recent SDW plus orbital ordering calculation (38) which shows a similar FS topology to our observation including the symmetry properties still does not contain any nesting. Therefore, the very straight FS pieces we observed appear well beyond the results of the structure and/or spin anisotropy of the system. Instead, we will argue that these strongly nested pieces of FS are consistent with the unidirectional nematic-like electronic nanostructures observed in Chuang *et al.*'s STM studies (10), with the mechanism for these also helping to drive the straight pieces of FS.

In fig. 4(b1), besides the FS nesting with the nesting vector $\vec{q}_1$ which is formed by the γ3 bands, we also introduce two other nesting vectors $\vec{q}_2$ (from the γ4 bands) and $\vec{q}_3$ (from the γ3 and γ4 bands). In fig. 2, we have shown that band γ3 has minimal $k_z$ variation while band γ4 does not have a Fermi crossing in the Γ plane, so only nesting vector $\vec{q}_1$ will persist through the whole momentum space and is the dominant one in the system. Such unidirectional straight parallel FS segments clearly indicate the one-dimensional nature of the electronic structure of the system. In our observation, the nesting vector $\vec{q}_1$=0.33(-π/a, π/a) may correspond to a unidirectional nanostructure with dimension approximately $6a_{Fe-Fe}$ in the real space. We have also made similar measurements on $Ca(Fe_{1-x}Co_x)_2As_2$ (x=3.5%) (not shown) which gives a smaller FS nesting vector $\vec{q}_1'$=0.24(-π/a, π/a), in-line with expectations from a semi-rigid shift of the chemical potential with doping. These two nesting vectors as a function of doping are plotted in fig. 1(a) by the black squares. We also plot in fig. 1(a) the inverse periodicity observed in the STM experiments (red square) which was $8a_{Fe-Fe}$ (10), finding that the STM results fall exactly in-line with that of the ARPES vector.

One can expect that by increasing the doping level, the nesting vector will keep getting smaller while at a specific point this incommensurate FS nesting will disappear due to the absence of the Fermi crossings of the hole-like bands. In real space, this means that the unidirectional nanostructure will no longer exist and the system will evolve into a new phase without the nematic property. The doping at which the FS nesting disappears would be a quantum critical point in the phase diagram. As shown by the dashed line in fig. 1(a), the simplest linear extrapolation indicates that this quantum critical point may occur near the end the superconducting dome. This therefore may point to a relationship between the quantum electronic liquid crystalline phases and the superconducting phase.

While the connection between the nanoscale nematic-like features observed in the STM experiments and the FS nesting observed here is solid, this does not fully explain the origin of either of these features, as the FS nesting is not expected in the electronic structure calculations. A tendency for the creation of nematic charge stripes is expected within certain theoretical models (4), and a feedback effect may be possible – the tendency to form nematic-like charge structures would be enhanced by the FS nesting tendencies, while these also may help drive the FS to have stronger nesting tendencies. Regardless, understanding the driving forces behind these unique structures is expected to be a hot topic in the coming years.

**Figure Legends**

Fig. 1 Phase diagram and asymmetric FS topology.

(a) Schematic phase diagram of the "122" pnictide system as a function of doping concentration (x) adopted from Ref. (10). The structural ($T_S$), antiferromagnetic ($T_{AF}$), and superconducting ($T_c$) transitions are shown. The two black squares indicate the size of the incommensurate nesting vector at 0 and 3.5% doping obtained from ARPES and the red square indicates the size of the nesting vector at 3% doing obtained from STM (10). The black arrowed dashed line indicates the simplest linear fitting based on these three points, which extrapolate to a possible quantum critical point in the overdoped regime. The inset of 1(a) shows the schematic of in-plane crystal structure of $CaFe_2As_2$. In the AFM-orthorhombic phase, the directions of the spins are shown as blue arrows on top of the Fe atoms. The solid red and green square boxes indicate the in-plane unit cell for the non-magnetic tetragonal state and the AFM-orthorhombic state, respectively. (b) The measured FS at 20K obtained by integrating spectra within an energy window of $E_F$ ±5 meV. Blue and green squares show the standard hole-like and electron-like Fermi surfaces, respectively, separated by the AF nesting vector $\vec{q}_0$. Here we focus more on the detail inside each Fermi surface pocket, and the smaller nesting vector $\vec{q}_1$ which matches the nematicity observed in STM (10). (c-g) Intensity maps near the zone center at different binding energies from $E_F$ to 80meV. All data were taken with 80eV photons.

Fig. 2 Incommensurate FS nesting and asymmetric $k_z$ dispersion.
(a,b) FS's taken with 80eV and 99eV photons at T=20K. The incommensurate FS nesting vector $\vec{q}_1$ is labeled on both plots. (c-f) ARPES intensity maps along the blue and yellow lines in panels (a) and (b). The Fermi crossings are indicated by the yellow and blue arrows on the plots. (g,h) $k_z$ dispersions of the Fermi crossings along the blue and yellow lines of (a) and (b). The dashed lines are guides for the eyes of the $k_z$ dispersions while the blue solid lines indicate the persistence of the nesting vector $\vec{q}_1$ in momentum space.

Fig. 3 Orbital determinations from polarization dependent measurements.

(a1-c1) Schematics of the three different ARPES experimental configurations. Symmetries with respect to the gray-shaded mirror plane are shown. (a2-c2) FS's, and (a3-c3) intensity plots along the high symmetry directions indicated by the white dashed lines in (a2)-(c2), respectively. (a4)-(c4) Dispersions along the high symmetry cuts obtained from (a3)-(b3), with allowable symmetries determined by the experimental geometries indicated. All data were taken with 80eV photons at T=20K.

Fig. 4 Comparison between (a,b) the experimental data and (c) theoretical calculations. Experimental orbital symmetry information is color-coded according to the polarization-dependent experiments (fig 3). We dashed the lines when there are multiple states that are consistent with the polarization data. The theory has been optimized for best agreement with the experiment by choosing a magnetic moment of $0.19\mu_B$, the bands with the minimal weight due the structure factor are removed from the image for better comparison, and the curves of (c2) have been renormalized by an overall factor of 2.5. The agreement with experiment is improved by shifting the yz bands up about 0.1 eV (arrows), which is indicative of orbital ordering.

39. The authors thank David Singh and Igor Mazin for helpful discussions. This work was supported by the Division of Materials Science and Engineering, Basic Energy Sciences, U.S. Department of Energy (U.S.D.O.E) grants DE-FG02-03ER46066, AC03-76SF00098 and DE-FG02-07ER46352 with supplemental support from the U.S. NSF grant DMR 0706657, and benefited from the allocation of supercomputer time at NERSC and Northeastern University's Advanced Scientific Computation Center (ASCC). The Advanced Light Source is supported by the Director, Office of Science (U.S.D.O.E.) under Contract No. DE-AC02-05CH11231. The work at Los Alamos National Laboratory was performed under the auspices of the U.S.D.O.E.

Fig.1

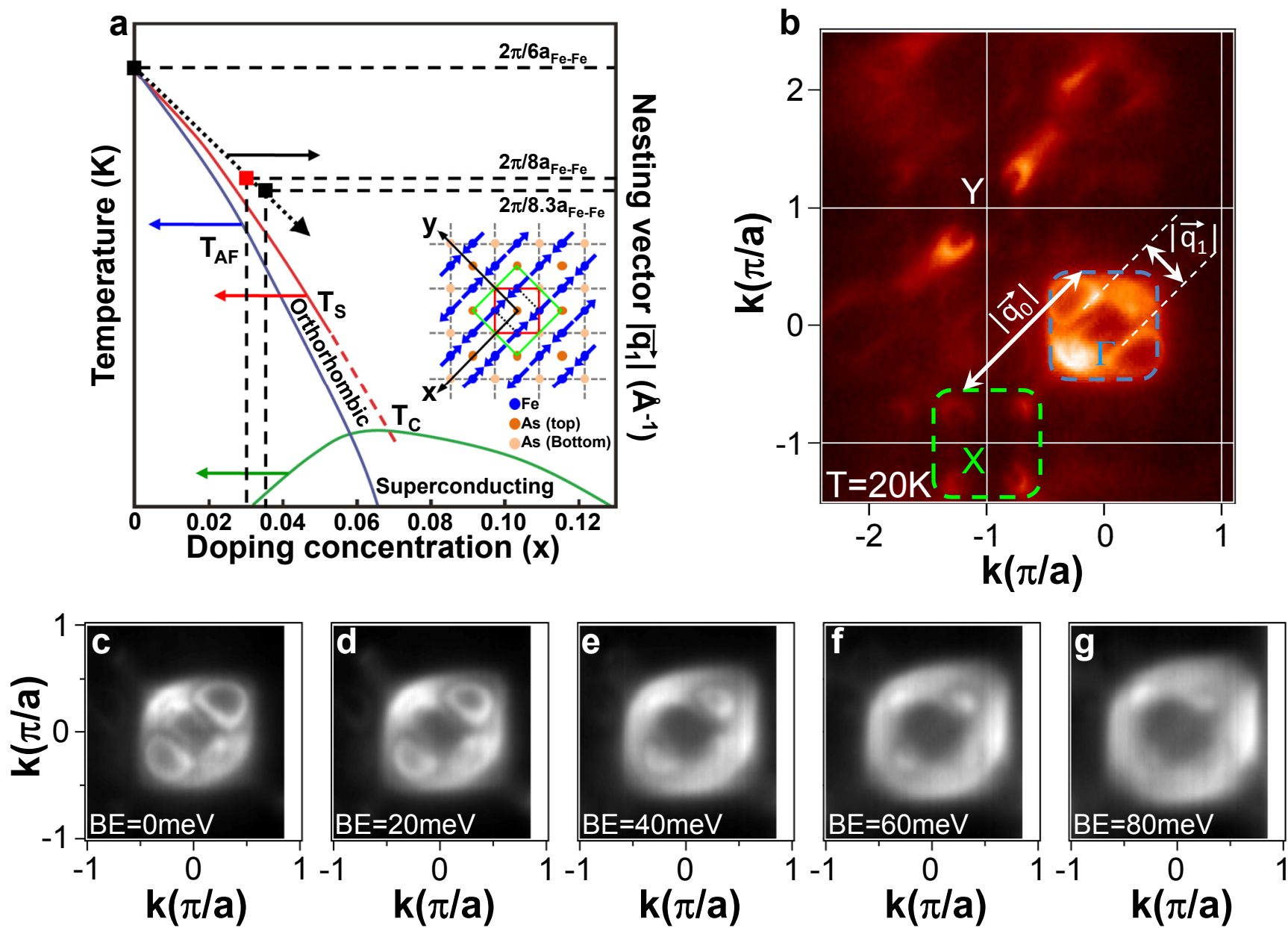

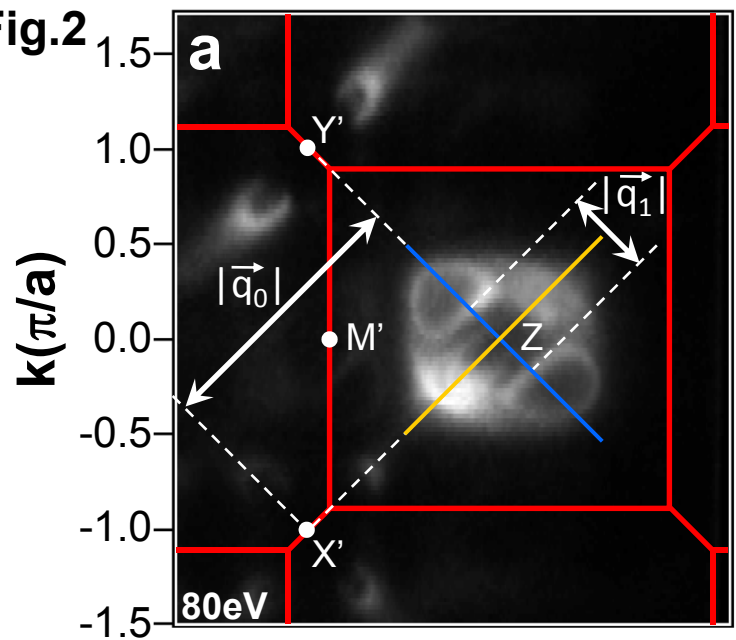
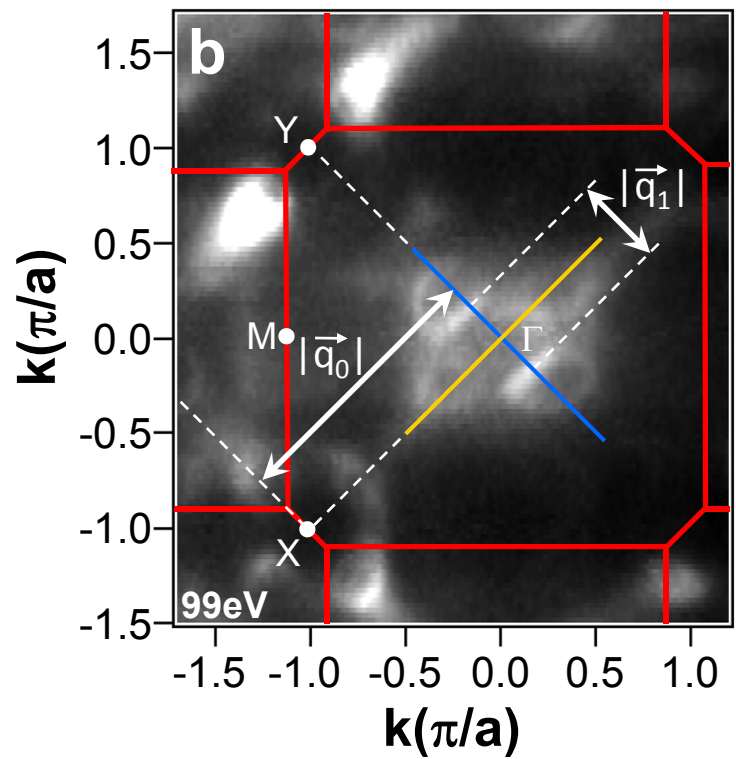
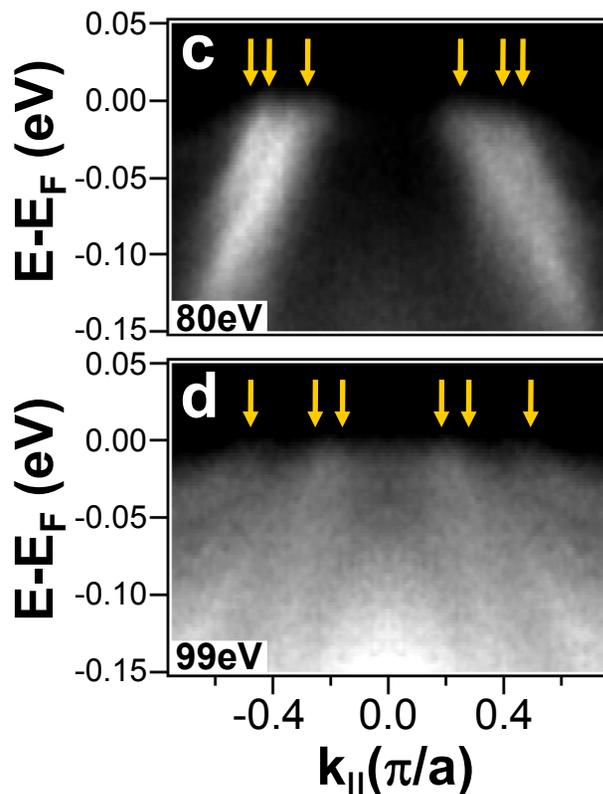
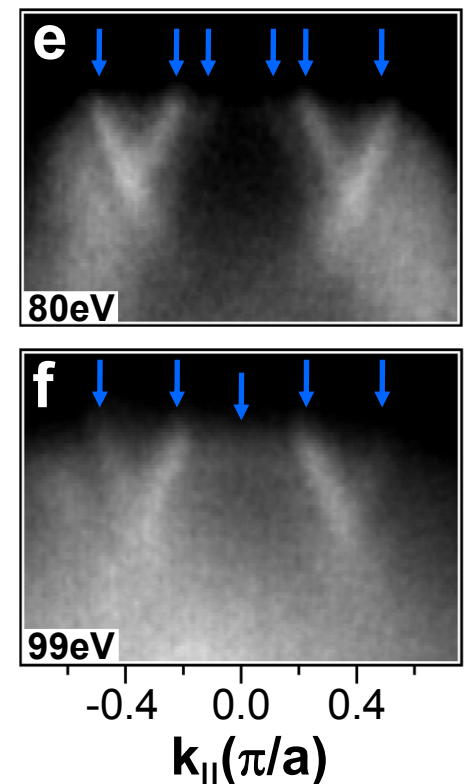
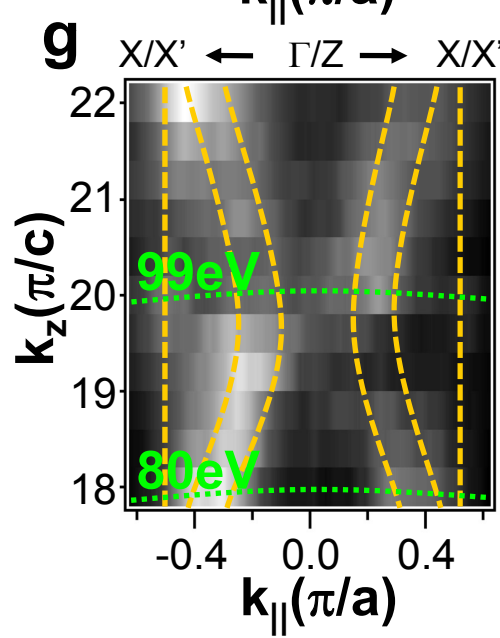
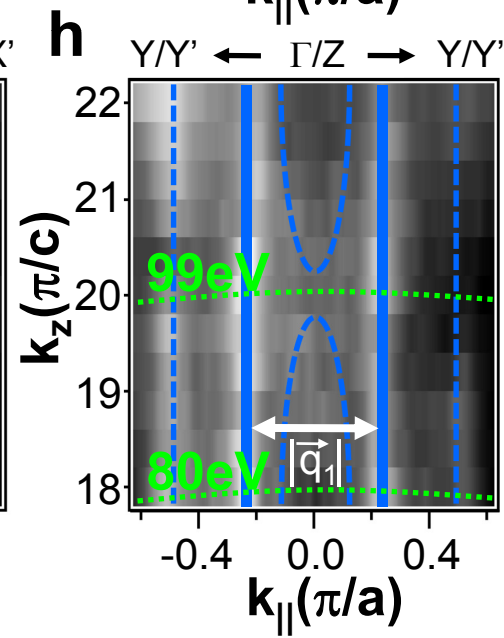

Fig.2

Fig.3

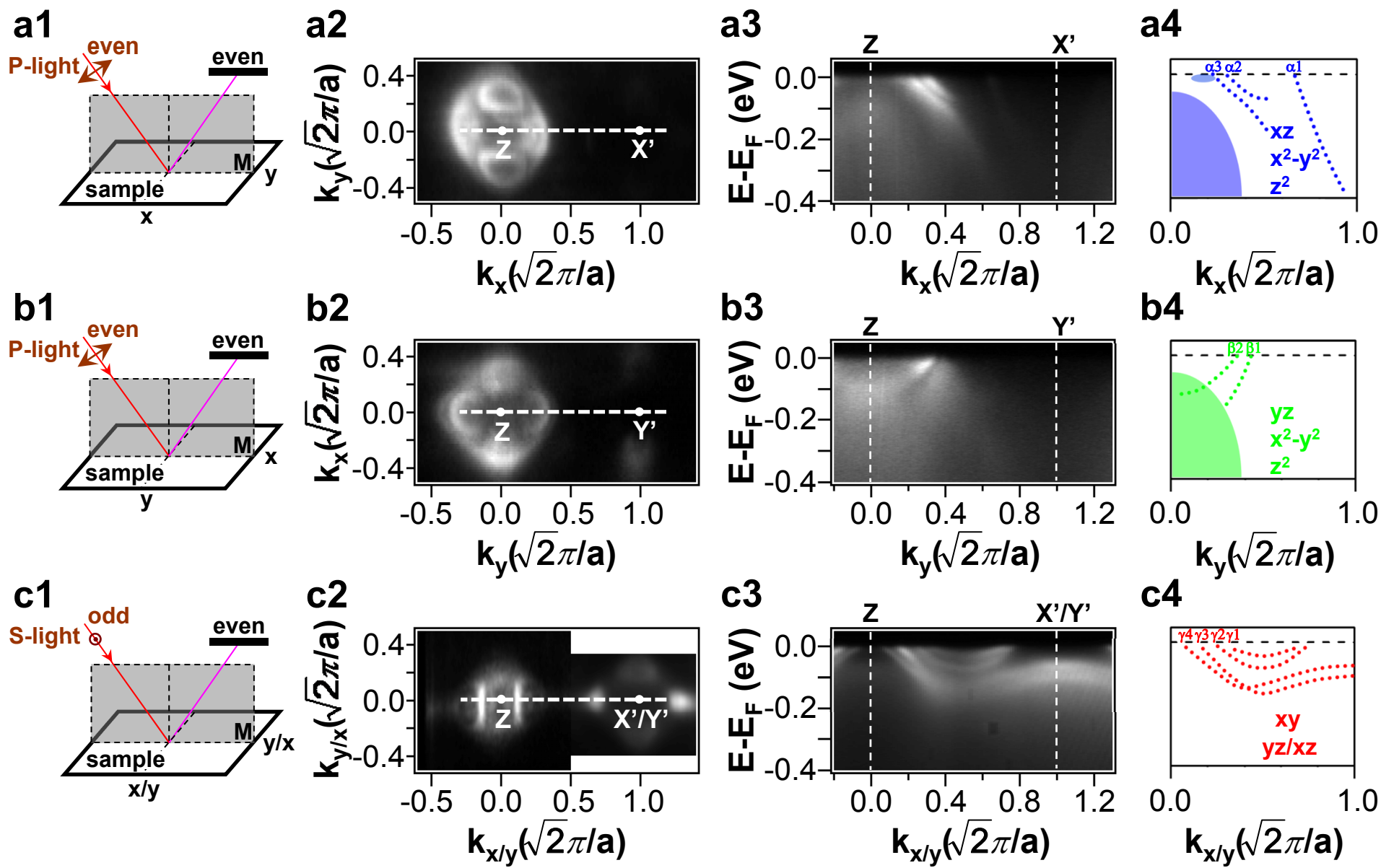

Fig.4

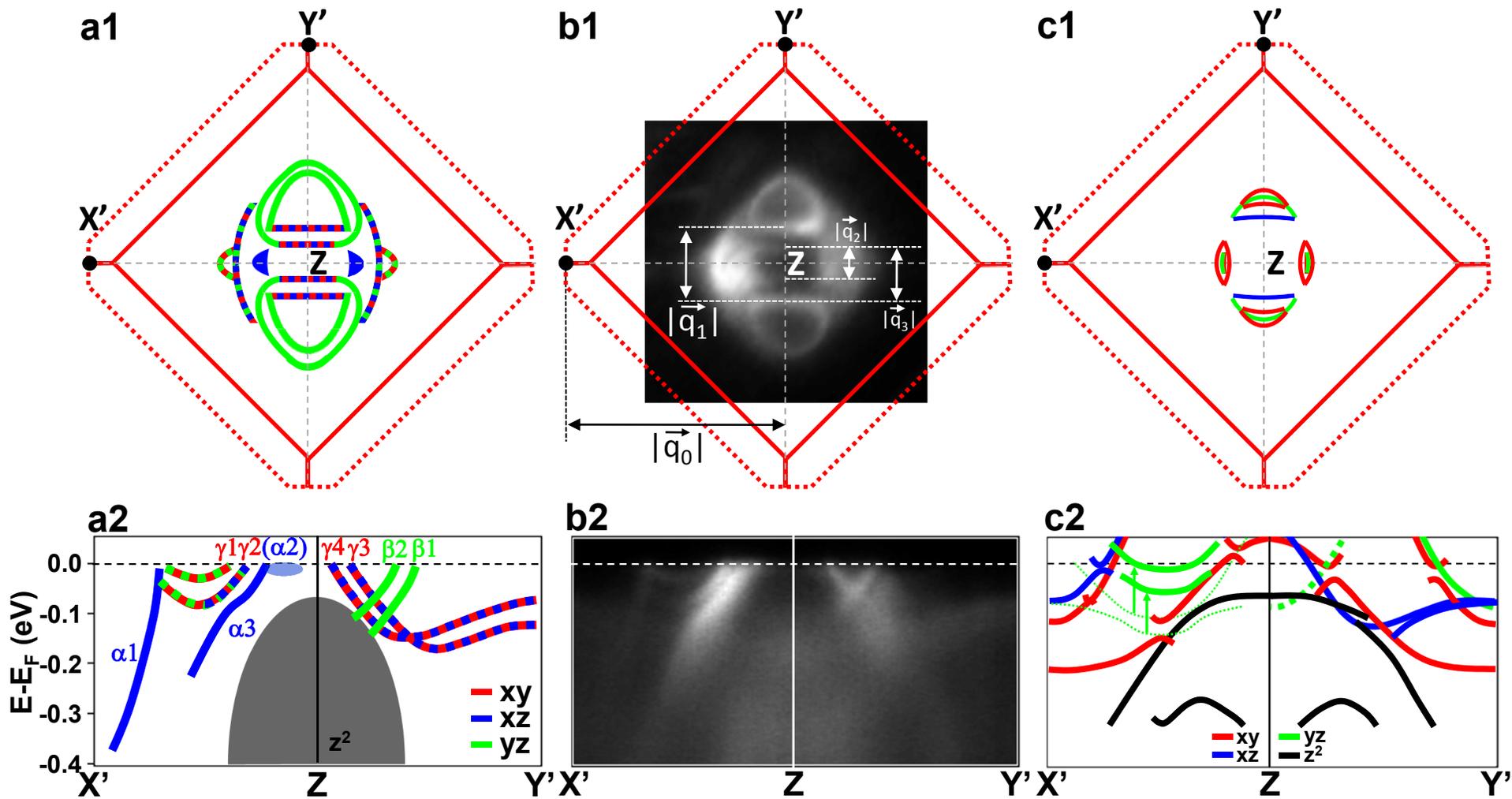